\documentstyle[prl,aps]{revtex}

\begin{document}
\twocolumn
\title
{
Transport in two dimensional periodic magnetic fields
}

\author
{
Junji Yoshida, Tomi  Ohtsuki and Yoshiyuki Ono$^{1}$
}

\address
{
Department of Physics, Sophia University,
Kioicho 7-1, Chiyoda-ku, Tokyo 102 \\
$^1$ Department of Physics, Toho University,
Miyama 3-3-1, Funabashi-shi, Chiba 274 
}

\maketitle

\begin{abstract}
{
Ballistic transport properties in a two dimensional electron gas are studied 
numerically, where magnetic fields are perpendicular to the plane of 
two dimensional electron systems   
and periodically modulated both in $x$ and $y$ directions.
We show that there are three types of trajectories of classical electron 
motions in this system; chaotic, pinned and runaway trajectories.
It is found that the runaway trajectories can explain the peaks of 
magnetoresistance as a function of external magnetic fields, 
which is believed to be related to the commensurability effect between 
the classical cyclotron diameter and the period of magnetic modulation. 
The similarity with and difference from the results in the antidot lattice  
are discussed.
}
\end{abstract}

\section{Introduction}

High mobility two dimensional electron gases formed in GaAs/AlGaAs 
heterostructures allow ballistic  motion of electrons over distances of 
several $\mu$m at low temperatures\cite{A1}.
Ballistic system is realized at low temperatures, low electron densities  
and low impurity concentrations.
Such samples are used in many experiments to study various novel transport 
properties  in low dimensional mesoscopic semiconductors, and a lot of 
interesting phenomena have been discovered.
One of such mesoscopic systems is a two dimensional electron gas (2DEG) 
in a periodic antidot array.

When the sample's typical geometrical scale, such as the distance between antidots,  
is larger than the Fermi wave length, the dynamics of a wave 
packet approaches the classical limit.
Semi-classical model explains some experimental results in this system, such as 
the quenching of the Hall resistance in low fields\cite{A2,A3}, 
and the oscillatory field-dependence of the 
magnetoresistance in an antidot array \cite{A4,A5,A6}.
In the latter, the potential is periodically modulated, and the 
commensurability of the period of potential and the cyclotron radius plays an 
important role.
Variety of interesting features appear in the diagonal and Hall resistances 
due to this commensurability effect.
Most of the features are now well understood in terms of classical orbits.
For example, the peak structure of magnetoresistance is explained by the
existence of runaway orbits.\cite{A10,A11}
Detailed analyses of the Hall resistivity have made clear the step-like
structure of the Hall resistivity accompanied by the peak of the
diagonal resistivity,\cite{IA} and the negative and vanishing Hall resistivity
in low field regime.\cite{nagao}
The quantum transport is also studied extensively \cite{AUIN} to clarify the
Altshuler-Aronov-Spivak effect \cite{NA} and the Anderson localization \cite{UA}
in antidot systems.

Then, the question arises. 
Can we observe similar effect when the magnetic fields instead of potentials 
are modulated ?
The answer is yes.
Many peculiar features are observed experimentally when the magnetic field 
is modulated one dimensionally\cite{A7,A71} 
and two dimensionally\cite{A8,A9}.

In this paper, we numerically calculate the magnetoresistance as well as 
the Hall resistance in the presence of two dimensionally modulated 
magnetic fields.
A classical billiard model is adopted and use is made of the Kubo formula 
to obtain the transport coefficients.
Such a classical model is justified, since the experimentally observed 
temperature dependence 
of the transport coefficients is weak, suggesting that the quantum 
interference 
effects may be neglected.

In the next section, we explain the model in detail, 
and in the third section, we present the results.
The final section is devoted to summary and discussions.

\section{Model and Method}                

Besides a uniform magnetic field $B_{\rm U}$, 
the periodically modulated fields $B_{\rm M}$ are applied perpendicular to the 
plane of 2DEG in the model.
Experimentally, these modulated magnetic fields are formed by ferromagnetic metals or a type II superconductor put on the surface 
of the heterostructure.

We introduce dimensionless variables
\begin{equation}
\begin{array}{ccccc}
 \tilde{{\bf r}}&=&(\tilde{x},\tilde{y})&=&(x/a_{0},y/a_{0}) ,\nonumber \\
 \tilde{{\bf v}}&=&(\tilde{v}_{x},\tilde{v}_{y})&=&
(v_{x}/v_{\rm F},v_{y}/v_{\rm F}) ,  \nonumber \\ 
 \tilde{t}&=&t/\tau_{0} & & , 
 \end{array}
\label{E1}
\end{equation}
where $a_{0}$ is the basic lattice constant of the artificial structure of $B_{\rm M}$,
$v_{\rm F}$  is the Fermi velocity, and $\tau_{0}=a_{0}/v_{\rm F}$.
 We scale the magnetic field by $B_{0}=2mv_{\rm F}/ea_{0}$, 
where $m$ is the effective mass of electron, and $e$ is the electron charge. 
Thus $B=B_{0}$ corresponds to a cyclotron radius of $a_{0}/2$.
We ignore inelastic scattering mechanism in this system, and therefore 
the classical electron velocity 
$\tilde{v}=\sqrt{\tilde{v}_{x}^{2} +\tilde{v}_{y}^{2}}$ is conserved.

The equations of motion is given by

\begin{eqnarray}
 \frac{{\rm d} \tilde{{\bf r}}}{{\rm d} \tilde{t}}&=& \tilde{{\bf v}} , \\ 
 \frac{{\rm d} \tilde{{\bf v}}}{{\rm d} \tilde{t}}&=& -2\tilde{{\bf v}}\times
 \tilde{{\bf B}} ({\bf r}),
 \label{E2}
\end{eqnarray}   
where $\tilde{B}({\bf r})$ is a total magnetic field perpendicular 
to the plane of the 2DEG and scaled by $B_{0}$.
It is a sum of the external uniform and periodically modulated magnetic field,
\begin{equation}
  \tilde{B}({\bf r})=\tilde{B}_{\rm U}+\tilde{B}_{\rm M}({\bf r}) .
\label{E3} 
\end{equation}

We use two types of modulated magnetic fields $B_{\rm M}$.
In one model (type I), $B_{\rm M}$ is finite and constant only in a finite 
region of radius $r_{\rm M}$ and zero otherwise (Fig. 1(a)).
In the other model (type II), 
a smooth space variation is assumed for the modulated field (Fig. 1(b)). 
Whether we take type I model or type II depends on how we realize the 
periodically modulated fields experimentally.

In the first case (type I), the magnetic field $B$ is constant and changes 
discontinuously in the designed region which exists at intervals. 
In the unit cell whose center is at the origin, it is expressed as 


\begin{equation}
B_{\rm M}(\bf r)=
\left\{
\begin{array}{ll}
 B_{\rm M}^{0} & ,  |{\bf r}|<r_{\rm M}    \\
 0    &  ,  |{\bf r}|>r_{\rm M}    .
\end{array}
\right.
\label{E4} 
\end{equation}

In the second case (type II), we use smooth function for $B_{\rm M}$.
\begin{equation}
  B_{\rm M}(x,y)=B_{\rm M}^{0}[\cos(c_{x}x)\cos(c_{y}y)]^{\beta}  ,
\label{E5} 
\end{equation}
where $\beta$ controls the steepness of the modulated magnetic field.
The factor $B_{\rm M}^{0}$ in eq.(\ref{E5}) determines the maximum value of 
the modulated magnetic field, and $c_{i} (i=x,y)$ controls the period of 
modulation.
We consider here only the case where $c_{x}=c_{y}=\pi/a_{0}$.

We apply the classical linear response theory\cite{A10} to calculate the conductivity tensor, 
and $\sigma_{ij}$ is given by
\begin{equation}
 \sigma_{ij}\propto 
\int_{0}^{\infty}e^{-\tilde{t}/\tilde{\tau}}  
\langle\tilde{v}_{i}(\tilde{t}_{0}+\tilde{t})
\tilde{v}_{j}(\tilde{t}_{0})\rangle {\rm d}\tilde{t} .
\label{E6}   
\end{equation}
where
$\langle \tilde{v}_{i}(\tilde{t}_{0}+\tilde{t})
\tilde{v}_{j}(\tilde{t}_{0})\rangle$ is the 
velocity correlation averaged over initial condition.
Here $\tilde{\tau}$ is the average time scaled by $\tau_{0}$ 
between two impurity scatterings in this system.
Impurity scatterings are not taken into account in the calculation of 
the equation of motion eq.(\ref{E2}), 
assuming that the impurity scattering effect and the classical motion are 
separable as expressed in eq.(\ref{E6}). 
A similar assumption was adopted in ref.\cite{A10} to obtain 
magnetoresistance oscillations in an antidot array. 
In this paper we chose $\tilde{\tau}=\tau/\tau_{0}\approx 4$.
We compute 
$\langle \tilde{v}_{i}(\tilde{t}_{0}+\tilde{t})
\tilde{v}_{j}(\tilde{t}_{0})\rangle$ by numerically
integrating eq.(\ref{E6}) to obtain $\sigma_{ij}$.
In the actual simulation, ten thousands of random initial conditions are 
chosen for the average, and the resulting error
(standard deviation) of the data
points is typically a few percent.
From the conductivity tensor, the magnetoresistance $\rho_{xx}$
and the Hall resistance $\rho_{\rm H}$ are expressed as
\[
 \rho_{xx}=\sigma_{xx}/(\sigma_{xx}^{2} +\sigma_{xy}^{2}) ,
\]
 \begin{equation}
 \rho_{\rm H}=-\sigma_{xy}/(\sigma_{xx}^{2} +\sigma_{xy}^{2}) .
\label{E7} 
\end{equation}

\section{Results}      

In Fig. 2, we show three typical trajectories found through a numerical 
simulation of the classical electron motion in periodic modulated 
magnetic systems (type I,II), chaotic, pinned and runaway 
trajectories\cite{A10,A11}.
The chaotic trajectory represents an ordinary diffusion mechanism.
Most trajectories belong to this type of the chaotic trajectory which appear 
in  all magnetic fields we examined.
The pinned trajectory is a non-diffusive trajectory which exists only near 
commensurability external magnetic fields ($\pm B_{0}$).
It is confined around a modulated magnetic field for a long time, 
and does not contribute to the diffusion.
The runaway trajectory skips between the periodic modulation  along $x$ 
or $y$ direction.
This trajectory appears suddenly from chaotic or pinned trajectories 
as the cyclotron diameter becomes commensurate with the lattice 
constant. 
It does not persist for a very long time.
 It is a chaotic trajectory in a strict sense.

Figure 3 shows a typical result for the longitudinal resistance $\rho_{xx}$ 
and the Hall resistance $\rho_{\rm H}$ 
in a type I model as a function of the externally applied field $B_{\rm U}$.
The period of the magnetic modulation is $a_{0}$,
the radius of the modulated field region $r_{\rm M}$ is $0.2a_{0}$,
and the strength of the modulated magnetic field $B_{\rm M}^{0}$ is $10B_{0}$.
Resistances are scaled by $\rho_{0}$ which is the resistance of the system 
without the field modulation and without the applied external field. 
First, we report the behavior of the longitudinal resistance 
(Fig. 3, the solid line).
We observe two features.  
One is that the peaks of the longitudinal resistance clearly appear 
near the  commensurability magnetic fields ($\pm B_{0}$).
The other is the asymmetry of magnetoresistance about zero field.
These structures are characteristic in the periodically modulated 
magnetic fields.

We then change the strength of the modulated field $B_{\rm M}$.
The commensurability peak decreases with the decrease of a modulated field
$B_{\rm M}$ and disappears at low field (Fig. 4).
The peaks disappear when the cyclotron diameter in the modulated region is 
larger than the diameter of the modulated region.
The magnetoresistance is influenced by the shape and strength of the modulation. 

In case of smooth modulated model (type II), the structure of the 
longitudinal resistance is similar to type I (Fig. 5).
However, the peak of the smooth modulated model shifts to a lower magnetic 
field when we use the same value for $B_{\rm M}^{0}$ as in type I.
When the steepness parameter $\beta$ increases, the peak position 
approaches to 
the one in the type I model.

The Hall resistance $\rho_{\rm H}$ also shows peculiar features.
In Fig. 3, we observe that the magnetoresistance peaks are accompanied 
by the abrupt increase of $\rho_{\rm H}$.
This is also observed in the type II model (Fig. 5).
Furthermore, in the type II model, $\rho_{\rm H}$ shows three plateaus  
between the sharp changes. 
There is the asymmetry of magnetoresistance about zero field.
Note that
the Hall resistance is still positive when $B_{\rm U}=0$.

The step-like structure of $\rho_{\rm H}$ in antidot systems
has been analyzed in
detail and explained by Ishizaka and Ando.\cite{IA}
They have introduced a hopping model to simplify the
classical motion of electrons, and  have classified the trajectories
(called {\it components})
by the successive center coordinates of the antidots which
scatter the electron.
The step of $\rho_{\rm H}$ has been shown to arise from the
cancelation of the contributions from the right going components and
the left going ones.
As seen from Fig.  2, the chaotic motions of the electrons in
modulated magnetic fields can also be mapped onto such
simplified components,
though the scattering mechanism is completely different.
Therefore, the step-like structure accompanied by the
peak of the diagonal resistivity in this system is also
explained according to the picture.
The non-vanishing value of $\rho_{\rm H}$ at $B_{\rm U}=0$ 
will be discussed in the next section.

We have also examined the other lattice structure, namely the triangular 
lattice. In this case, runaway trajectory skips along 
$(1,0), (1/2,\sqrt{3}/2)$ or $(-1/2,\sqrt{3}/2)$ direction, 
and the peaks in the resistivity again appear at the commensurability fields.

\section{Summary and Discussion}                     

The above results that peaks of the longitudinal resistance occur near 
the commensurability magnetic fields suggests that pinned or runaway 
trajectories contributes to the peaks.
The pinned trajectory at commensurate magnetic field can not explain 
the difference of magnetoresistance peak positions between positive and 
negative external field, nor the disappearance of the commensurability 
peaks at low modulated field,
since the pinned trajectory is not influenced by the direction 
of the external field and the strength of the modulation. 
An asymmetry of magnetoresistance about zero field and its dependence on 
strength of modulated magnetic field are explained only by the runaway
trajectory.
This means that runaway trajectories shown in Fig. 2 are essential to 
explain the simulation.

In the type I model, the rough estimate of the magnetic field that a runaway 
trajectory appears is given by 
\begin{equation}
B_{\rm U}(B_{\rm M}^{0})=\frac{B_{0}}{1+B_{0}/B_{\rm M}^{0}}
\;\;\;\;\; ( B_{\rm U}>0, B_{\rm M}^{0}>0 ),
\label{E81}
\end{equation}
\begin{equation}
B_{\rm U}(B_{\rm M}^{0})=-\frac{B_{0}}{1-B_{0}/B_{\rm M}^{0}}
\;\;\;\;\; ( B_{\rm U}<0, B_{\rm M}^{0}>0 ).
\label{E82}
\end{equation}

Figure 6(a) and (b) show typical runaway trajectories at fields given by eqs.(\ref{E81}) and (\ref{E82}), respectively. 
When $B_{\rm U}$ and $B_{\rm M}$ are in the same direction, 
$2R_{\rm c}=a_{0}+2R_{\rm M}$ is the condition that a runaway trajectory appears,
where $R_{\rm c}$ is the cyclotron radius in the absence of modulation,
and $R_{\rm M}$ is the cyclotron radius in the modulated region.
When $B_{\rm U}$ and $B_{\rm M}$ are in the opposite directions, 
the condition changes to $2R_{\rm c}=a_{0}-2R_{\rm M}$.
The peak positions of the magnetoresistance obtained in 
the simulations are in good agreement with the field values given by  
eqs.(\ref{E81}) and (\ref{E82}), as indicated in Fig. 3 by arrows.
In smoothly modulated magnetic field, electron motion is bended more strongly 
than in the absence of modulation.
In this case, the runaway trajectory appears in the lower field than the field 
expected from eqs.(\ref{E81}) and (\ref{E82}).
This is the reason why peaks for the type II model shift to a lower field.

 As stated in the previous section, $\rho_{\rm H}$ remains to be
positive even when the external field $B_{\rm U}$ is vanishing.
To explain this behavior,
we have introduced the effective magnetic fields. 
We introduce an averaged magnetic field $B_{\rm eff}$ 
felt by electron as 
\begin{equation}
B_{\rm eff}=\lim_{T\rightarrow\infty}\frac{1}{T}
\int_{0}^{T}B({\bf r}(t))dt , 
\label{E11}
\end{equation}
where ${\bf r}(t)$ is the position of the electron.
The obtained $B_{\rm eff}$ looks almost linearly dependent on $B_{\rm U}$ 
and is well described by
\begin{equation}
B_{\rm eff}/B_{0}=\alpha\tilde{B}_{\rm U}+\gamma .
\label{E12} 
\end{equation}
This effectively shifts the point of $B_{\rm eff}=0$  
to a negative value of $B_{\rm U}$ (Fig. 7).
Here $\alpha$ and $\gamma$ are constant and  
the values are summarized in table I.
In type I model, $\alpha$ increases and becomes almost equal to unity 
as $B_{\rm M}^{0}$ increases. 
This is because the electron cannot stay in the modulated field region 
for long when $B_{\rm M}^{0}$ is large.

This idea of the effective field qualitatively explains the positive 
value of $\rho_{\rm H}$ at $B_{\rm U}=0$, since 
$B_{\rm eff}>0$ even if $B_{\rm U}=0$.
However, $B_{\rm eff}$ remains positive even when $\rho_{\rm H}=0$. 
One might think that such vanishing of $\rho_{\rm H}$ for positive 
$B_{\rm eff}$ 
can be explained by the vanishing of the Hall resistivity\cite{A1,A2}, 
but we do not observe the plateau or non-monotonic behavior 
around $\rho_{\rm H}=0$ which is typical for the vanishing $\rho_{\rm H}$. 
The fact that $\rho_{\rm H}(B_{\rm U}=0)>0$ and
$\rho_{\rm H}(B_{\rm eff}=0)<0$ 
means that the non-uniform magnetic fields act  as scatterers as well as
magnetic fields bending the electron trajectories and giving rise 
to the Hall effect.

To summarize, we have studied the transport properties in two dimensional 
periodic magnetic fields by numerical simulation using classical model 
and explained the mechanism of magnetoresistance peaks 
accompanied by sharp changes of the Hall resistance by the 
runaway trajectory.
Runaway trajectories are found to play a very important role in 
periodically modulated 
magnetic fields 
similarly as in the antidot array systems.

In real experiments performed so far \cite{A7,A71,A8,A9},
attaching ferromagnets to produce modulated magnetic fields
necessarily introduces a strain induced electric potential as well,
while in our simulation, we have assumed that the
effect of magnetic field modulation is much stronger than that of the
scalar potential modulation and have neglected its effect.
Furthermore, the field modulation is assumed to be stronger than
the externally applied field.
Such assumptions might be satisfied by introducing the modulated
field by attaching type II superconductor.\cite{geim}
Density modulated composite fermion systems with $\nu=1/2$\cite{HLR,KLKG,NF}
may also be a possible candidate.

\vspace{10pt}

{\bf Acknowledgements}

\vspace{10pt}

We would like to thank Dr. P. D. Ye and Dr. K. Tsukagoshi for 
valuable discussions.

\bigskip

\noindent
{\bf Figure captions}

Fig. 1 : Schematic view of the type I model (Fig. 1(a)) and type II (Fig. 1(b)). 
In case of type I, an external constant magnetic field $B_{\rm U}$ is applied 
perpendicular to the plane outside the region enclosed by circles  
and constant magnetic field $B_{\rm M}^{0}$ exists inside the circles.
In case of type II, The modulated magnetic field $B_{\rm M}$ is 
expressed by eq.(\ref{E5}) and changes smoothly.

\medskip
Fig. 2 : Typical three trajectories; pinned, chaos and runaway in 
the type I (Fig. 2(a)) and the type II (Fig. 2(b)) models.

\medskip
Fig. 3 : Longitudinal and Hall resistance in the type I model.
The  modulation of periodic magnetic field forms square lattice
($r_{\rm M}=0.2a_{0}, \tau=4a_{0}/v_{\rm F}$). 
Solid line is the longitudinal resistance, while the dotted line is the Hall 
resistance $\rho_{\rm H}$.
The magnetoresistance is normalized by the longitudinal resistance $\rho_{0}$ 
in zero field.
The modulated magnetic field $B_{\rm M}^{0}=10B_{0}$.
Typical standard deviations of the data points are indicated by
error bars.

\medskip
Fig. 4 : Magnetoresistance for various modulated magnetic field $B_{\rm M}^{0}$ 
in the type I model.
Solid line shows longitudinal resistance with $B_{\rm M}^{0}=10B_{0}$,  
while the dotted line is for $B_{\rm M}^{0}=5B_{0}$, 
and the dashed line for $2.5B_{0}$. 
Typical standard deviations of the data points are indicated by
error bars.
At low field, the peak structure disappears.

\medskip
Fig. 5 : Longitudinal and Hall resistance in smoothly modulated fields
(type II).
$\beta$ in eq.(\ref{E5}) is set to be 4.
The solid line shows longitudinal resistance and the dashed line shows
Hall resistance. 
The maximum magnetic field $B_{\rm M}^{0}=10B_{0}$.
Again,  the typical standard deviations of the data points are indicated by
error bars.

\medskip

Fig. 6 : Typical runaway trajectory given by eqs.(\ref{E81}) and (\ref{E82}).
In Fig. 6(a), a typical runaway trajectory where $B_{\rm U}$ and $B_{\rm M}$ 
are in the same direction is shown, while in Fig. 6(b),
a typical runaway trajectory where $B_{\rm U}$ and $B_{\rm M}$ are in
the opposite directions is shown.

\medskip

Fig. 7 : Effective magnetic fields $B_{\rm eff}$ vs. $B_{\rm U}$ 
in type I model with $B_{\rm M}^{0}=10B_{0}$.


\bigskip
\begin{center}
Table I  \\
\end{center}
The coefficients $\alpha$ and $\gamma$ in each modulation 
( $B_{\rm M}^{0}/B_{0}=2.5,5,10$ ). 
We choose $r_{\rm M}=0.2a_{0}$ in type I model and $\beta=4$ in type II.

\medskip
\begin{center}
\begin{tabular}{ccccc} \hline
 & \makebox[15mm]{\({\displaystyle  B_{\rm M}^{0} }\)}
 & \makebox[15mm]{\({\displaystyle 2.5B_{0} }\)}
 & \makebox[15mm]{\({\displaystyle 5B_{0} }\)}
 & \makebox[15mm]{\({\displaystyle 10B_{0} }\)} \\ 
\hline
type I   & \({\displaystyle  \alpha  }\) 
         & 0.88
         & 0.90
         & 0.94      \\
         & \({\displaystyle  \gamma  }\)
         & 0.29
         & 0.47
         & 0.58     \\
\hline      
type II  & \({\displaystyle  \alpha  }\) 
         & 1.00
         & 0.98
         & 0.86     \\
         & \({\displaystyle  \gamma  }\)
         & 0.34
         & 0.67
         & 0.99   \\
\hline
\end{tabular}
\end{center}


\begin{thebibliography}{99}
\bibitem{A1} For review article, see for example, C. W. J. Beenakker and 
H. van Houten: Solid State Phys. {\bf 44} (1991) 1.
\bibitem{A2} M. L. Roukes: Phys. Rev. Lett. {\bf 59} (1987) 3011.
\bibitem{A3} C. W. J. Beenakker and H. van Houten: 
Phys. Rev. Lett. {\bf 63} (1989) 1857.
\bibitem{A4} D. Weiss, M. L. Roukes, A. Menschig, P. Grambow, K. von Klitzing 
and G. Weimann: Phys. Rev. Lett. {\bf 66},(1991) 2790;
for review article, see for example,
D. Weiss, G. L\"utjering and K. Richter:Chaos, Solitons and Fractals {\bf 8} (1997) 1337.
\bibitem{A5} K. Tsukagoshi, S. Wakayama, K. Oto, S. Takaoka 
and K. Murase: Phys. Rev. B {\bf 52} (1995) 8344.
\bibitem{A6} K. Tsukagoshi, T. Nagao, M. Haraguchi, S. Takaoka, 
K. Murase and K.Gamo: J. Phys. Soc. Jpn. {\bf 65} (1996) 1914.
\bibitem{A10} R. Fleischmann, T. Geisel and R. Ketzmerick: 
Phys. Rev. Lett. {\bf 68} (1992) 1367.
\bibitem{A11} {\'E}. M. Baskin, G. M. Gusev, Z. D. Kvon, 
A. G. Pogosov and M. V. {\'E}ntin: JETP Lett. {\bf 55} (1992) 678.
\bibitem{IA} S. Ishizaka and T. Ando: Phys. Rev. B {\bf 55} (1997) 16331.
\bibitem{nagao} T. Nagao: J. Phys. Soc. Jpn. {\bf 66} (1997) 3183.
\bibitem{AUIN} T. Ando, S. Uryu, S. Ishizaka and T. Nakanishi:
Chaos, Solitons and Fractals {\bf 8} (1997) 1057.
\bibitem{NA} T. Nakanishi and T. Ando: Phys. Rev. B{\bf 54} (1996) 8021.
\bibitem{UA} S. Uryu and T. Ando: in the proceeding of the
12th international conference on electronic properties of
two dimensional systems (EP2DS-12), Physica B
{\bf 249-251} (1998) 308;
S. Uryu: Ph.D. thesis, University of Tokyo (1997).
\bibitem{A7} S. Izawa, S. Katsumoto, A. Endo and Y. Iye: 
J. Phys. Soc. Jpn. {\bf 64} (1995) 706.
\bibitem{A71} M. Kato, A. Endo and Y. Iye: 
J. Phys. Soc. Jpn. {\bf 66} (1997) 3178.
\bibitem{A8} P.D.Ye, D.Weiss, K. von Klitzing, k. Eberl, H. Nickel: 
Appl. Phys. Lett. {\bf 67} (1995).
\bibitem{A9} P. D. Ye, D. Weiss, R. R. Gerhardts, M.Seeger, 
K. von Klitzing, K. Eberl and H.Nickel: Phys. Rev. Lett. {\bf 74} (1995) 3013.
\bibitem{geim} A.K. Geim, S. J. Bending and I.V. Grigorieva:
Phys. Rev. Lett. {\bf 69} (1992) 2252.
\bibitem{HLR} B. I. Halperin, P.A. Lee and N. Read:
Phys. Rev. B{\bf 47} (1993) 7312.
\bibitem{KLKG} S.A. Kivelson {\it et al.}:
Phys. Rev. B{\bf 55} (1997) 15552.
\bibitem{NF} N. Nagaosa and H. Fukuyama: to appear in
J. Phys. Soc. Jpn.


\end{thebibliography}
\end{document}